\documentclass[%
preprint,
 amsmath,amssymb,
 aps,
prb,
]{revtex4-1}

\usepackage{graphicx,subfigure,epsfig}
\usepackage{dcolumn}
\usepackage{bm}

\begin{document}

\preprint{APS/123-QED}

\title{Nanostructures for very broadband or multi-frequency transition from wave beams to a subwavelength light distributions}

\author{O.~Luukkonen}
\email{olli.luukkonen@tkk.fi}
\affiliation{%
 Department of Radio Science and Engineering/SMARAD CoE,
 Aalto University, P.O. 13000, 00076 AALTO, Finland.
 }%

\author{C.~Simovski}
\email{konstantin.simovski@aalto.fi}
\affiliation{%
 Department of Radio Science and Engineering/SMARAD CoE,
 Aalto University, P.O. 13000, 00076 AALTO, Finland.
}%
\author{J.~Pniewski}%
\email{jpniewski@igf.fuw.edu.pl}
\affiliation{%
 Faculty of Physics, University of Warsaw, Pasteura 7, 02-093 Warszawa,
Poland
 }%

\date{\today}

\begin{abstract}
In this paper we suggest and theoretically study a tapered
plasmonic nanostructure which connects the incident wave beam with
a subwavelength spatial region where the field is locally enhanced
in a broad frequency range or for different operation frequencies.
This spatial region has a frequency stable location near the
contour of the tapered structure. This results from a special
waveguide mode which can also exist in the tapered structure. We
foresee many possible applications for our structure from
prospective near-field scanning optical microscopes to
interconnects between conventional optical waveguides and
prospective optical nanocircuits.\end{abstract}

\pacs{Valid PACS appear here}
\maketitle


\section{Introduction}

An actual problem for many branches of the modern optics is the
concentration of the light energy initially transported by a
conventional wave beam into a subwavelength spatial region. The
problem is not new and there is a whole body of literature devoted
to so-called subwavelength light concentrators which is difficult
to refer in this paper. In most of these works the objective is to
achieve very high resolution or very large local field enhancement
at a certain frequency. In works on plasmonic nanostructures
designed to enhance of thin-film solar cells (see, {\itshape e.g.}, Ref.~1) 
the subwavelength field concentration is achieved
together with efficient transmission of the light energy into this
region (photovoltaic layer of nanometer thickness). However, as a
rule, this subwavelength field concentration is a narrow-band
effect. Rarely, it is achievable with the same nanostructure
within 2-3 very narrow resonance bands (see, {\itshape e.g.}, Ref. 2) 

Most sub-wavelength light concentrators such as a solid metal tip
with nanoscale curvature of the apex or a nanoantenna operate in
the regime when the surrounding space is also illuminated. On the
contrary, tapered plasmonic nanostructures, we have unified under the
name of metamaterial nanoptips (MMNT),\cite{NT1,NT2,NT3} concentrate the light energy (in our
simulations) in a needed sub-wavelength region so that the surrounding space is not illuminated. This way the
parasitic excitation of the objects located outside the selected region with the incident wave beam with is avoided. Structures
capable of realizing this regime are also not new in the nano-optics. They
are actually being used in some schemes of field-enhanced
nanosensing, especially of the field-enhanced fluorescence,\cite{FEF} of the field-enhanced Raman spectroscopy,\cite{FES}
and also in field-enhanced microscopy \cite{FEM}. Beside these
applications, structures which realize such a regime could be used
for interconnections between conventional optical information
systems and prospective sub-wavelength devices. Sub-wavelength
devices which need to be connected with usual optical waveguides
are plasmonic nanowaveguides which have developed fast since the
past decade,\cite{P1,P2} optical nanocircuits,\cite{Engheta} and
active all-optical nanodevices (nanoswitches, nanotransistors,
nanodiodes and optical memory cells \cite{P3}). In all of these cases
the general problem can be formulated as follows: One needs to
transmit the optical signal from a conventional waveguide to a
needed nano-objects whose locations are known while minimizing the
cross-coupling to other nano-objects. This problem formulation
recently led to the general idea of a specially shaped ({\itshape e.g.}
tapered) waveguide in which the usual wave beam formed by
propagating spatial harmonics would concentrate in a
sub-wavelength region outside the waveguide, for example, near the
end of the waveguide.\cite{Stockman} We complement this problem
formulation by a requirement of a broad frequency range or a
multi-frequency operation.

Notice that such a mode concentration was realized in some
conventional tips used in scanning near-field optical microscopes
(SNOM), namely in those being composed of a conically shaped
optical fiber covered with a metal film having a nanometer
aperture at the end. However, in such tips the hot spot is located
inside the cone which causes almost all the incident wave to
reflect from the end of the tip and the energy transmission
throughout the tip is very low. Attempts to use the extraordinary
transmission effect \cite{ET1,ET2} for such structures are known.
However, only a rather modest improvement of the transmittance was
reported in these works and even this modest improvement was
achieved by introduction of very complicated nanostructures, for
example nanocorrugations in the metal film of the narrowest part
of the nanocone.\cite{ETNT}

More promising results were theoretically demonstrated for our
MMNT. In Refs.~3 and 17 
tapered waveguides performed as metamaterial prisms (in Ref.~3) 
it comprised silver nanorods oriented along the prism edge, in Ref.~17 
it was formed by silver nanoplates parallel to both the edge and the waveguide
axis) efficiently transformed the incident plane wave illuminating
the base of the wedge into a light nanojet. The light nanojet is a
light beam keeping strongly sub-wavelength width ($\lambda/4$) up
to the distance $\lambda/2$ from the nanotip apex. In Ref.~4 
a pyramid filled with a metamaterial of plasmonic
nanospheres (also illuminated from the base) allowed a very small
hot spot (of the order $\lambda/100$) at the apex. This light
concentration was achieved in these works together with negligible
scattering and acceptable reflection losses.

Structures studied in the present paper possess important
advantages compared to the pyramidal nanotip reported in Ref.~4. 
First, they are multi-resonant and the overlapping of
resonances can result in a very broad operational frequency band.
Secondly, the spatial domain of the enhanced electric field is
nearly the same for all operation frequencies. The multi-frequency
operation and the frequency stability advantageously distinguish
our structure from substantial plasmonic nanoparticles matched to
a conventional optical waveguide in Ref.~19. 
We believe therefore that our MMNT reported will give a significant
development to ideas of Ref.~19 
and other relevant works published to the present time.

\section{Suggested structure and its operation}

\subsection{The main idea}

\begin{figure}[t!]
\centering
\subfigure[]{\epsfig{file=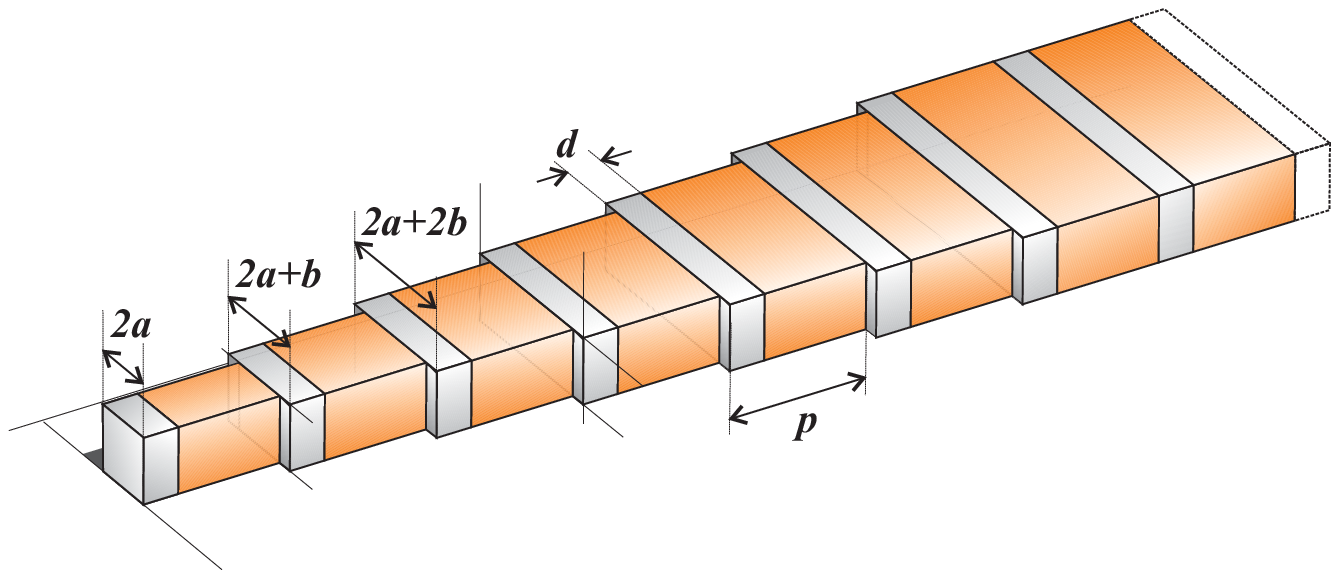,width=0.45\textwidth}}
\subfigure[]{\epsfig{file=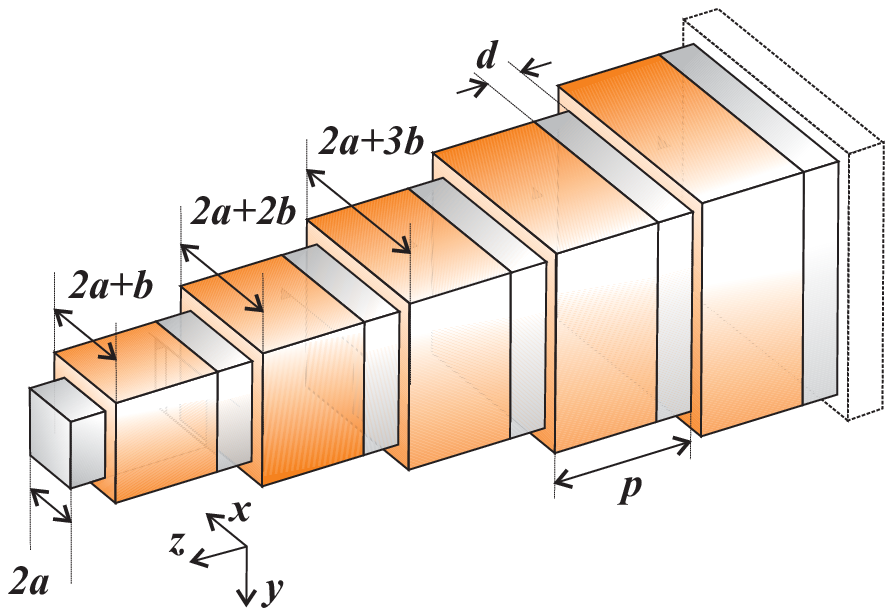, width=0.3\textwidth}}
\caption{(Color online.) Sketch of the proposed metamaterial structures for
broadband local field enhancement with stable position of the hot
spot: (a) -- planar structure; (b) -- pyramidal
structure.}\label{fig:1}
\end{figure}

The proposed structure comprises parallel rectangular
nanoparticles shaped as nanobars or as nanoplates of noble metal
(silver or gold) separated from one another with dielectric
spacers (also shaped as nanobars or nanoplates). Both variants are
shown in Fig.~\ref{fig:1}(a) and (b). In the first case the
structure width is much larger than its thickness. This MMNT can
be located at the dielectric substrate and excited by an
integrated ({\itshape e.g.} ribbon) dielectric waveguide having the same
substrate. A rather similar multi-resonance structure has been
known since 2008 and was called optical xylophone in Ref.~20. 
However, optical properties of the xylophone structure in which
rather wide metal nanobars are separated with narrow spacers are
different. The xylophone structure \cite{OX} would better
correspond to our Fig.~\ref{fig:1}(a) if we replace in this sketch
metal and dielectric nanobars. The optical xylophone located at
the dielectric interface was excited by an incident plane wave and
at a given operation frequency the local field was enhanced near
one metal nanoplate resonating at this frequency. Our structure
(planar MMNT) is excited by the wave beam impinging the structure
from the base of the structure, the field concentration holds in a
frequency tangle spatila region, the resonances of different
nanoplates overlap and form a very broad frequency range and all
this becomes possible due to novel plasmonic waveguide modes that
we called the edge modes.

Another MMNT studied below is a pyramid performed of stacked metal
nanoplates separated with dielectric spacers Fig.~\ref{fig:1}(b).
This nanostructure can be built at the end of an optical fiber. It
seems to be similar to the plasmonic waveguide studied in Ref.~21. 
However, due to different dimensions the operation of
our structure is different from that of the waveguide structure in Ref.~21. 
The purpose of the study in Ref.~21 
was the homogenization of the layered waveguide and the tapered variant of
the waveguide was studied in Ref.~21 
only in order to demonstrate that the tapering is possible without destruction of
the wave guidance. In Ref.~21 
plasmon resonances of nanoplates were not exploited. In our study we combine these resonances
together with the edge waveguide modes.

In both planar and pyramidal sets of nanoparticles the size of the
metal nanoparticles is gradually varied so that the plasmon
resonances of the structure cover a sufficient part of the visible
range.

\begin{figure}[t!]
\centering \epsfig{file=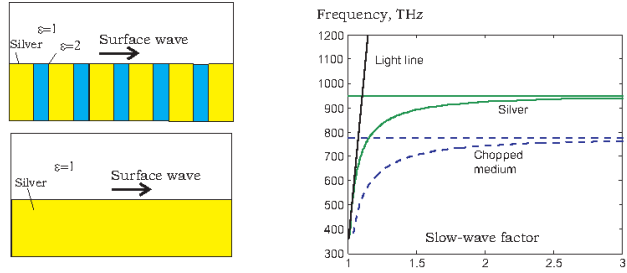,width=0.5\textwidth}
\caption{(Color online.) Left -- two structures for which the surface wave
dispersion curves were calculated (chopped interface and solid
silver interface). Right -- dispersion curves for these
structures. The dispersion for the chopped interface differs
sufficiently from the light line in the whole visible range.
}\label{fig11}
\end{figure}

The idea of the surface and edge modes arose from the analysis of
surface waves at the interface of the chopped medium with free
space. We call as the chopped medium a periodic layered
metal-dielectric structure with nanometer period. The sketch of
this medium is shown in the left panel of Fig.~\ref{fig11}. At
the interface of the chopped medium with free space the surface
wave excites which has a different dispersion than surface-plasmon
polariton (SPP) supported by the interface of the solid metal. The
dispersion of SPP for the solid silver sufficiently differs from
the light line only within the frequency interval $780-970$ THz.
This interval refers to the ultraviolet range. In the visible
frequency range the solid silver does not support surface waves.
The SPP resonance of the chopped medium occurs in the visible
range. In the example of the dispersion diagram presented in
Fig.~\ref{fig11} silver layers are of thickness $5$ nm and
dielectric layers with $\varepsilon=2.2$ are of thickness $10$ nm,
whereas the SPP resonance holds at $780$ THz. From $400$ to $750$
THz (the lower and upper bounds of the visible range) the
slow-wave factor is sufficiently larger than unity so that the
surface wave could be excited.

Fortunately it is possible to engineer such a wave with the
proposed structures because waves propagating along the sides of
the wedge- and pyramidal structures can be considered analogous to
the surface wave of the chopped medium. A simplified explanation
of this effect for a pyramidal structure in Fig.~\ref{fig:1}(b) is
as follows: At a given frequency the incident wave beam excites
one of the nanoplates, namely that resonating at this frequency. The
local field enhancement happens at the edges of the resonant plate
and this local field is a package of evanescent waves. One of
these evanescent spatial harmonics excites at the sides of the
pyramid a wave propagating from one nanoplate edge to another, {\itshape
i.e.}, we have a surface-like wave propagating along the sides of
the pyramid to its apex. The field of this wave is enhanced
compared to the incident wave at the surface of the pyramid and
decays in the surrounding space the further it is apart from this
surface. If the length of the nanostructure is properly chosen the
local field at the apex of the nanostructure will be be enhanced
more compared to the sides of the pyramid due to the constructive
interference of partial surface waves propagating at four sides of
the pyramid. The excitation of nanoplates which are some distant
away from the pyramid base is possible if the whole structure is
optically rather small and if, for instance, the thickness of
silver plates does not exceed $10-15$ nm. So thin silver plates
are nearly transparent in the visible.

The operation of the planar structure shown in Fig.~\ref{fig11}(b)
differs from the pyramidal case only by the dimensionality. The
similar edge wave propagating along the sides of the tapered
planar waveguide can be excited due to the same physical
mechanism. In this case the requirement of the small optical size
of the structure can be strongly softened compared to the
pyramidal structure. Slender metal nanobars reflect the incident
wave beam weakly. Therefore the waveguide mode of the input
waveguide can be matched with the mode of the tapered metamaterial
waveguide. We called the eigenwave responsible for the operation
of both planar and pyramidal structures the edge wave since its
electric field is mainly concentrated at the nanobar or nanoplate
edges. In order to optimize the tapered structure the edge mode
has been analyzed for non-tapered periodic waveguides.

\subsection{Dispersion of the edge mode}

\begin{figure}[t!]
\centering
\includegraphics[width=0.25\textwidth]{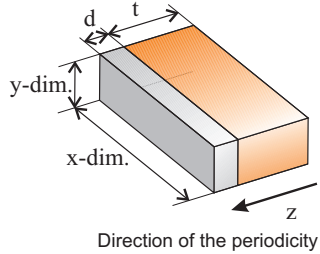}
\caption{(Color online.) A unit cell of the periodic plasmonic waveguide.}
\label{waveguidepic}
\end{figure}

\begin{figure}[t!]
\centering
\subfigure[]{\epsfig{file=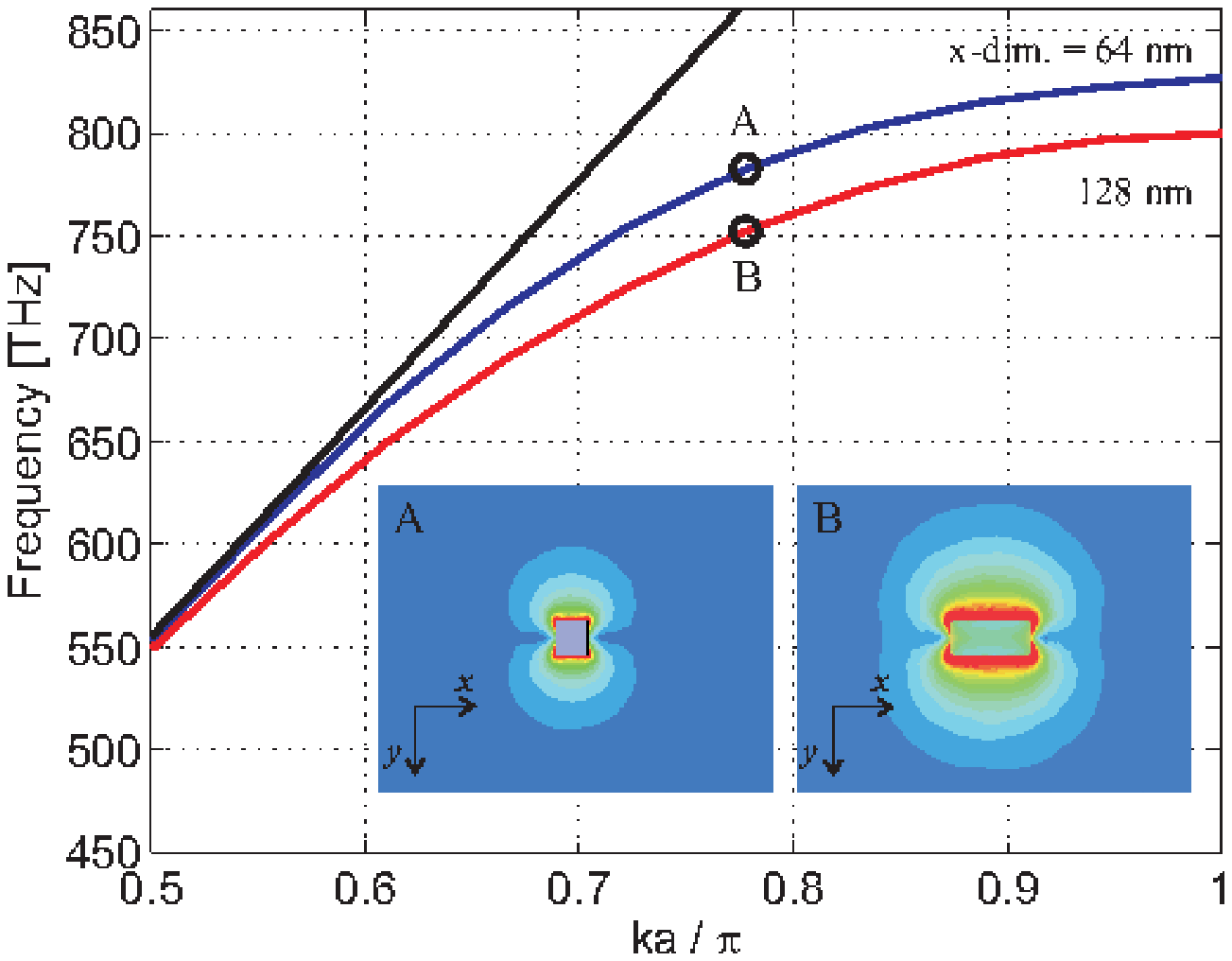,width=0.45\textwidth}}
\subfigure[]{\epsfig{file=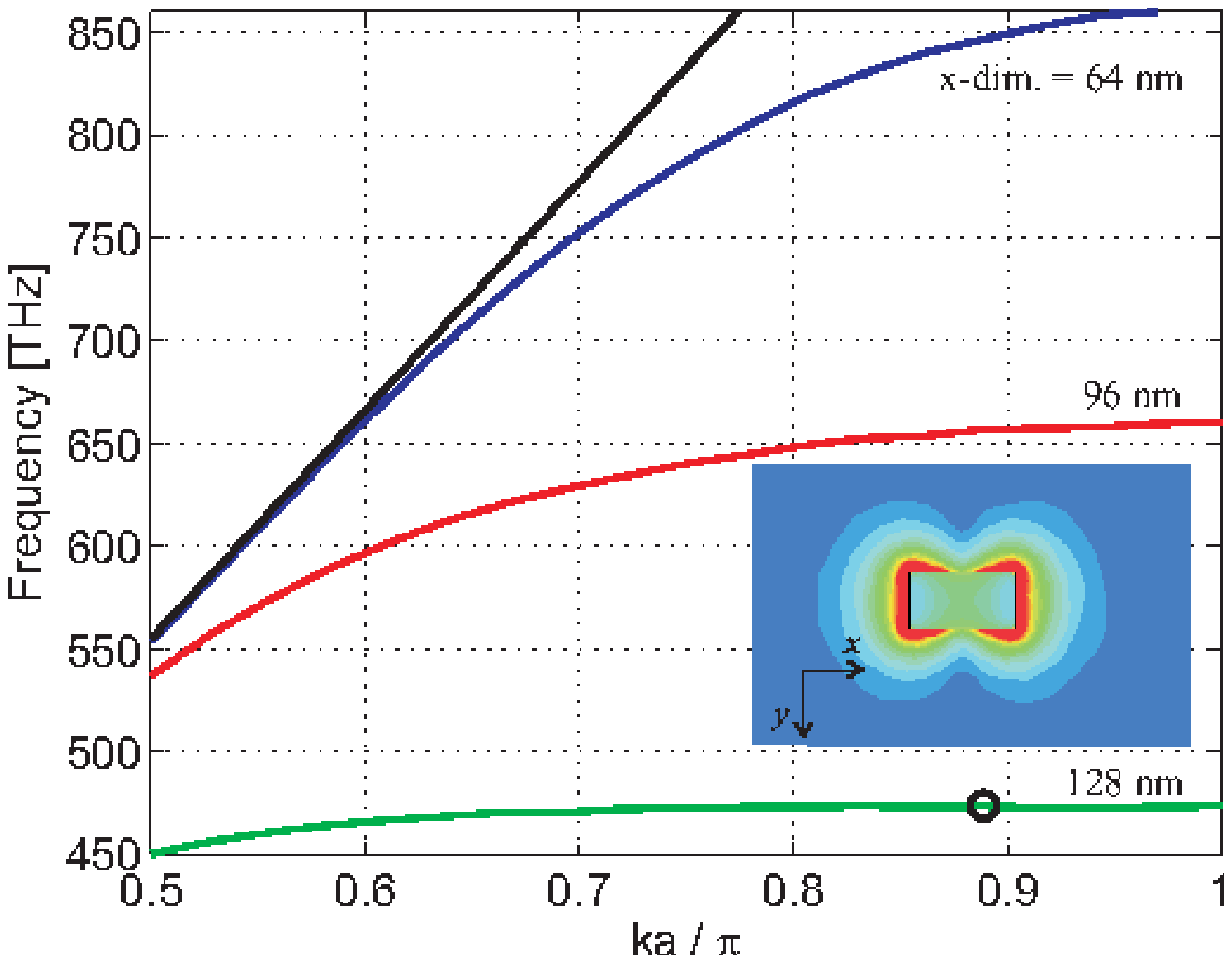,width=0.45\textwidth}}
\caption{(Color online.) Dispersion diagrams of the modes (a) -- on the horizontal
edges and (b) -- on the vertical edges for the plasmonic waveguide
with different $x-$dimensions. The other parameters read:
$y-$dimension is 70\,nm, $z-$dimension is 35\,nm, and $t=100$\,nm.
The period of the waveguide is 135\,nm. The amplitude distribution
of the electric field over the unit cell at the frequency denoted
by the black circle is shown in the insets.}
\label{dispersiondiagram}
\end{figure}

We numerically studied the dispersion properties of a planar
periodic plasmonic waveguide whose unit cell is shown in
Fig.~\ref{waveguidepic}. In this waveguide silver nanobars are
sandwiched in between dielectric nanoplates. The dispersion
characteristics of such a waveguide were studied using the HFSS
simulator whereas the complex permittivity of the silver
corresponded to the known experimental data \cite{JC} and the
permittivity of the dielectric was set equal to $2.2$. The
dispersion diagrams for different values of the $x-$dimensions
(keeping $y-$ and $z-$dimensions fixed) are shown in
Figs.~\ref{dispersiondiagram}(a) and (b). The amplitude of the
electric field of the different modes at different frequencies are
shown in the insets of these Figures. The edge modes shown in
Fig.~\ref{dispersiondiagram}(a) propagate on the horizontal side
of the waveguide whereas the modes depicted in
Fig.~\ref{dispersiondiagram}(b) propagate on the vertical side of
the waveguide.

We see that the modes propagating along the horizontal side of the
waveguide are less affected by the variation of the $x-$dimension
of the waveguide whereas the dispersion of the modes propagating
along the vertical side of the waveguide is clearly affected by
any change in the $x-$dimension of the waveguide. For our purposes
the mode which is less affected by the tapering of the waveguide
in the $x-$direction is more favorable. This makes possible the
tapering of the waveguide and a large operational frequency band.
The apex of the waveguide can be reduced to sub-wavelength
dimensions without sacrificing too much in the matching between
the wider waveguide sections. The operational band of the wedge is
determined by the frequency band over which the dispersion curve
of the mode on widest and narrowest waveguide sections overlap.
Clearly, we can see that for the vertical edge mode the
operational band hardly pushes to the optical regime. For the
horizontal edge mode on the other hand the operational band covers
the whole optical regime. Further, for both modes the most part of
the electric field energy is concentrated near the surface of
nanobars, outside their bulk. Therefore the energy absorption in
these modes is not so high.

\subsection{Plasmonic resonances of nanoparticles}

One can roughly estimate the resonance frequencies of an
individual nanobar (or nanoplate) replacing it by a slender (or
oblate) ellipsoid. For optically small ellipsoids one can use the
following condition for the internal electric field
$\mathbf{E}_{\rm i}$:\cite{Sihvola}

\begin{equation}
\mathbf{E}_{\rm i} = \frac{\varepsilon_{\rm e}}{\varepsilon_{\rm e} + N_i\left(\varepsilon_{\rm i} - \varepsilon_{\rm e}\right)} \mathbf{E}_{\rm e} \rightarrow \infty,
\end{equation}
where $\varepsilon_{\rm e}$ is the relative permittivity of the
surrounding media, $\varepsilon_{\rm i}$ is the relative
permittivity of the ellipsoid (in our case it is the complex
permittivity of silver), and $N_i$ is the depolarization factor
along $i=x,y,z$ coordinate axes. For the oblate
ellipsoids (spheroids) the depolarization factors read:\cite{Osborn}

\begin{equation}
N_{\rm z} = \frac{1 + e^2}{e^3}\left(e - \arctan\left(e\right)\right),
\end{equation}
\begin{equation}
N_{\rm x} = N_{\rm y} = \frac{1}{2}\left(1 - N_{\rm z}\right),
\end{equation}
where eccentricity $e = \sqrt{a^2/c^2 - 1}$.

The resonance frequency becomes then
\begin{equation}
f_i = \frac{1}{2\pi}\frac{\omega_{\rm p}}{\sqrt{1 - \varepsilon_{\rm e}\left(1 - N_i^{-1}\right)}} \label{res_freq}.
\end{equation}
In these estimations we neglect losses and the permittivity of
silver is assumed to obey the Drude dispersion
\begin{equation}
\varepsilon_{\rm i} = 1 - \frac{\omega_{\rm p}^2}{\omega^2},
\end{equation}
where $\omega_{\rm p}=\sqrt{N q^2/m \varepsilon_0}$ is the plasma
frequency, $N$ is the charger density, $q$ is the electron charge,
and $m$ is the electron mass. Naturally, the estimated resonance
frequencies calculated using the above equations are not quite the
ones we have for the rectangular nanoparticles, moreover those in
presence of dielectric spacers and those coupled with one another
through the waveguide mode. Nevertheless, this rough approximation
gave us a valid estimation of the resonance frequency range and of
the ratio between different resonance frequencies.

\section{Simulated performance of metamaterial structures}

\subsection{Planar nanotip}

\begin{figure*}[t!]
\centering
\mbox{
\subfigure[]{\epsfig{file=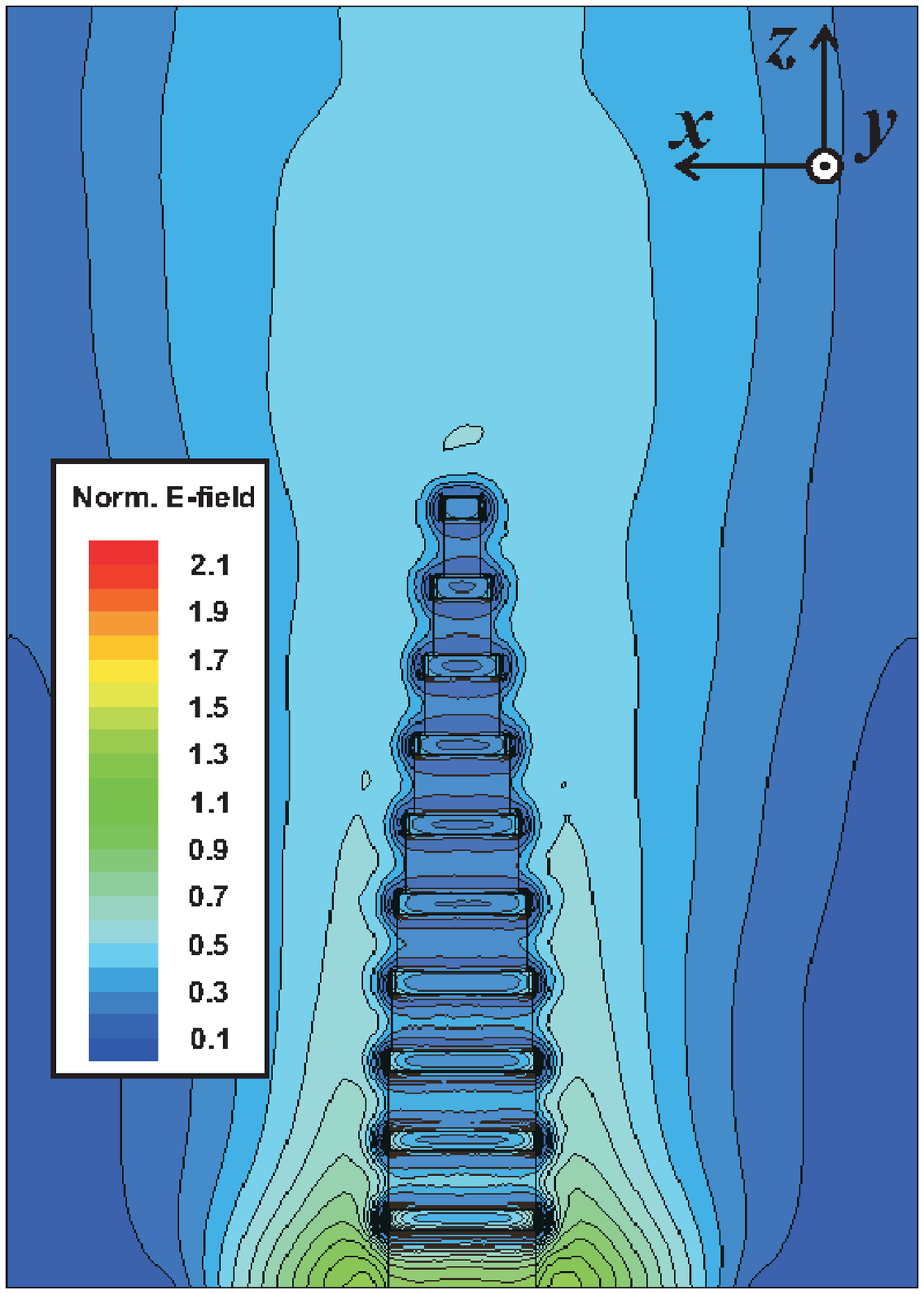,width=0.2\textwidth}}
\subfigure[]{\epsfig{file=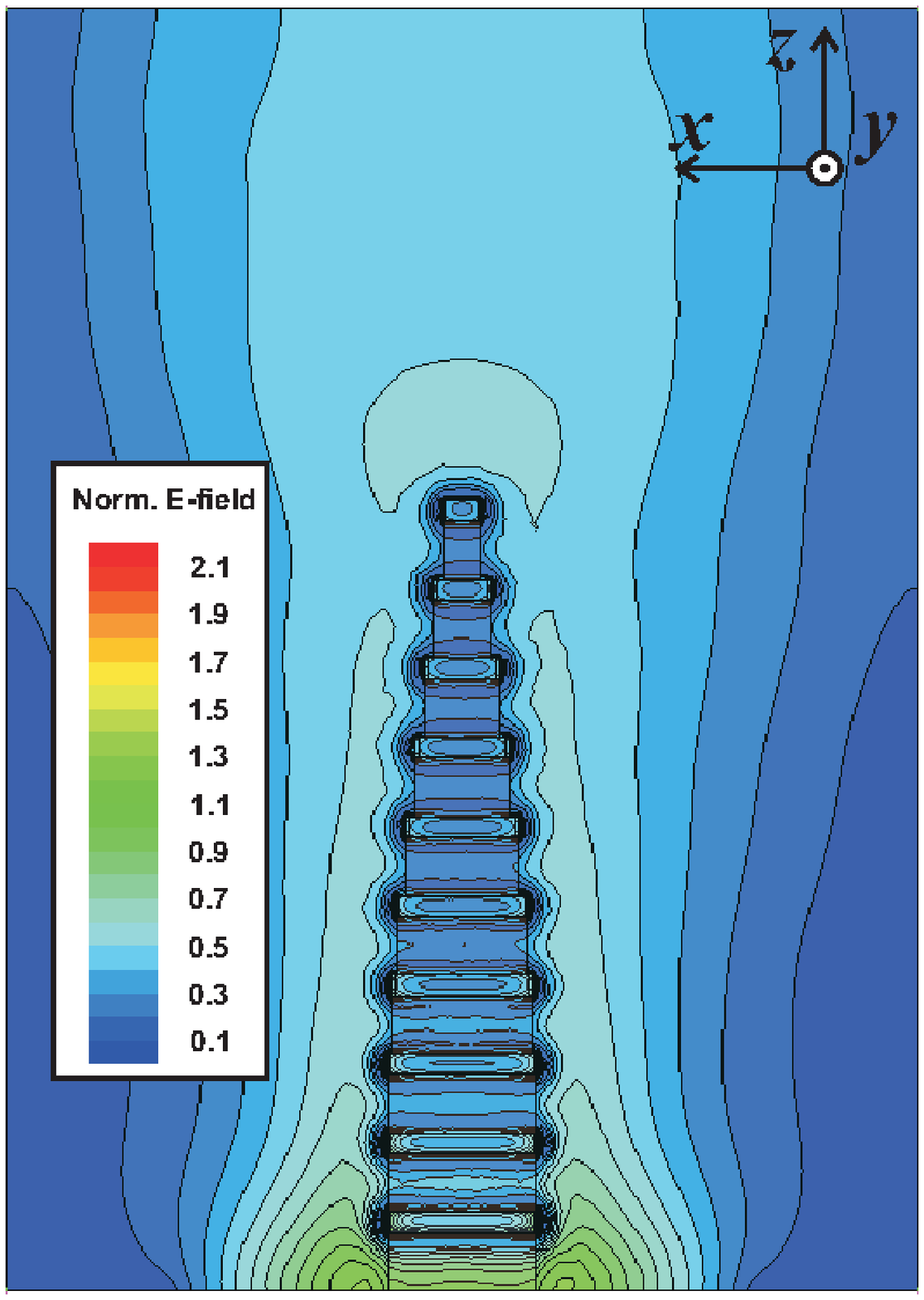,width=0.2\textwidth}}
\subfigure[]{\epsfig{file=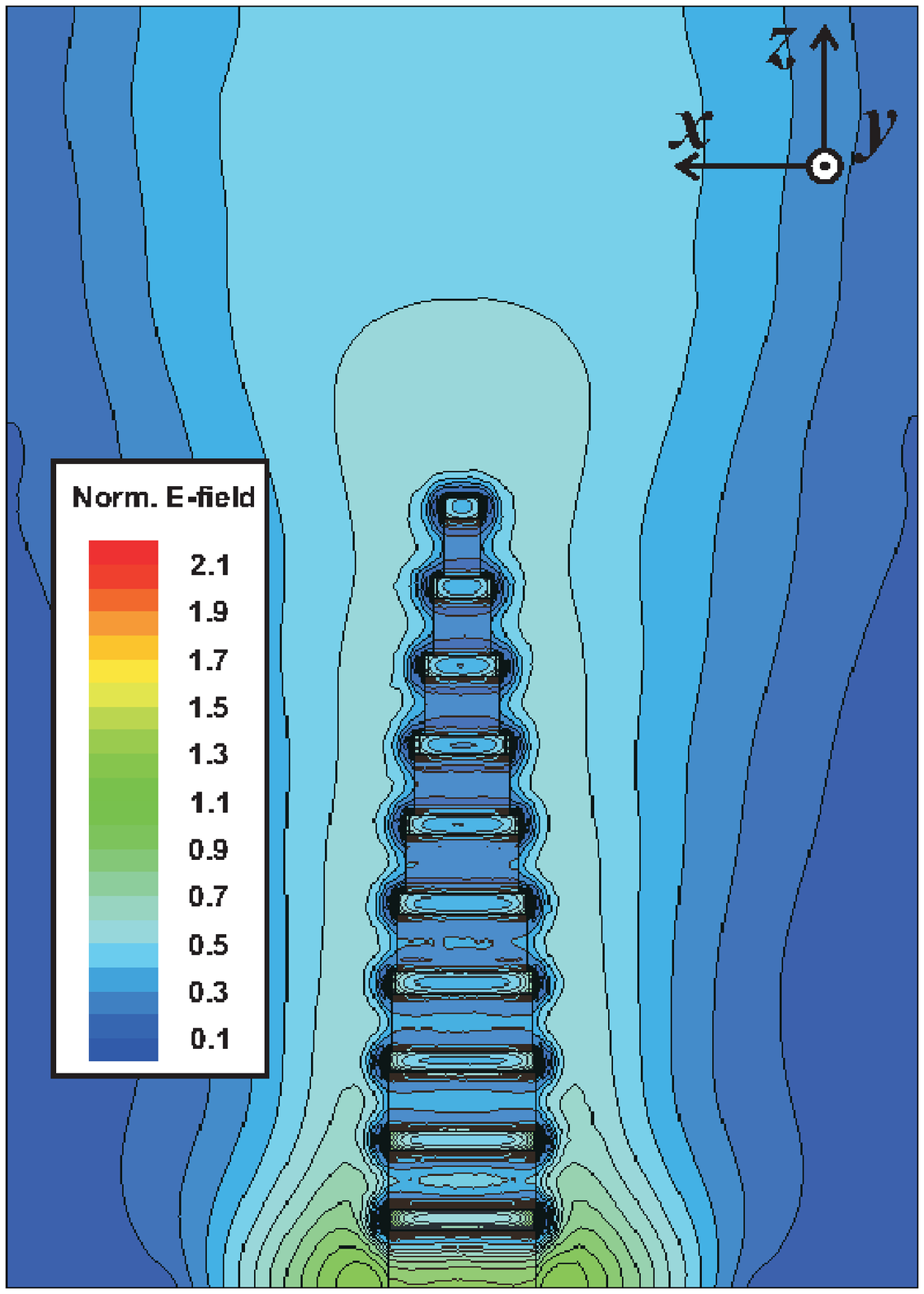,width=0.2\textwidth}}
\subfigure[]{\epsfig{file=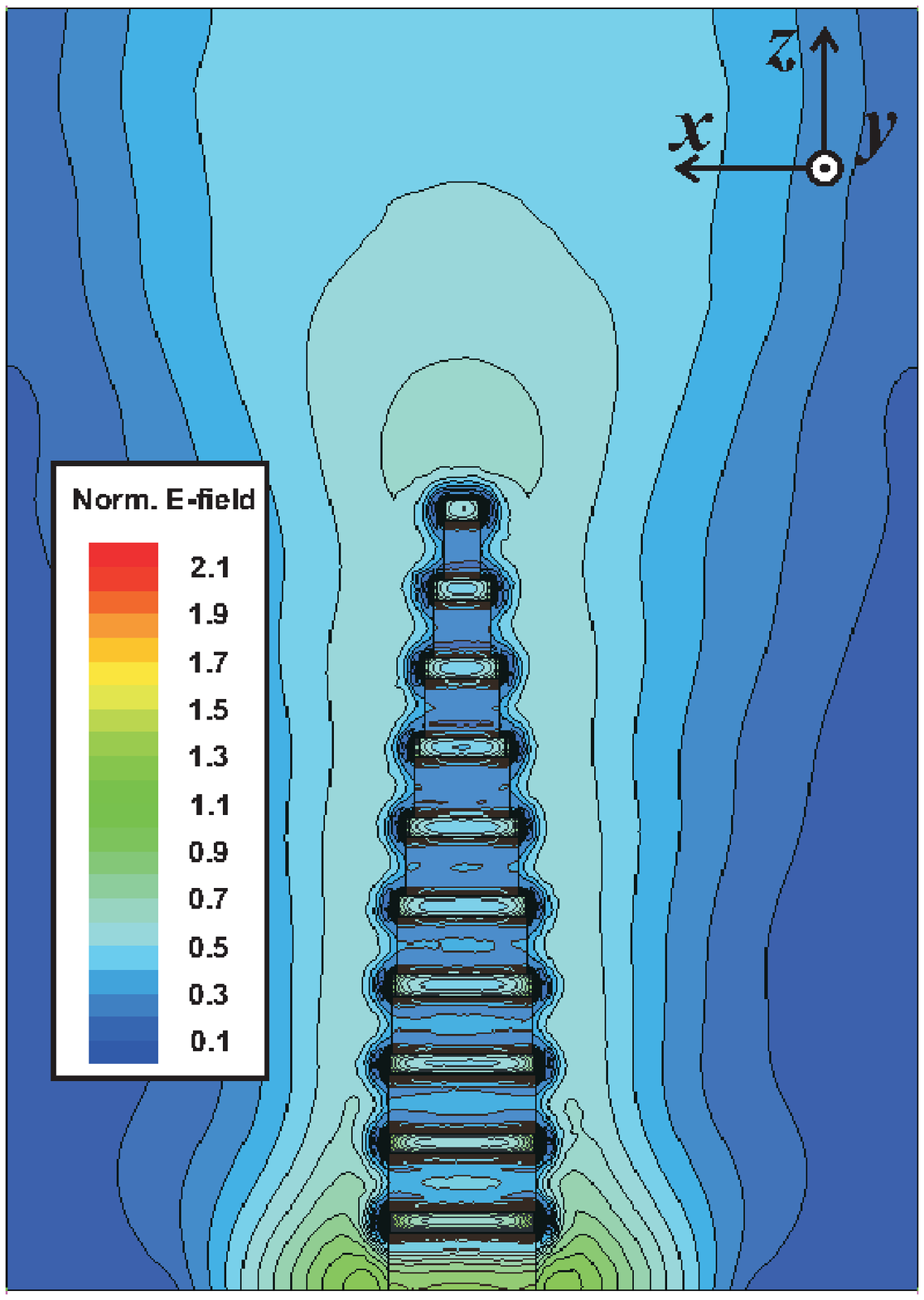,width=0.2\textwidth}}
\subfigure[]{\epsfig{file=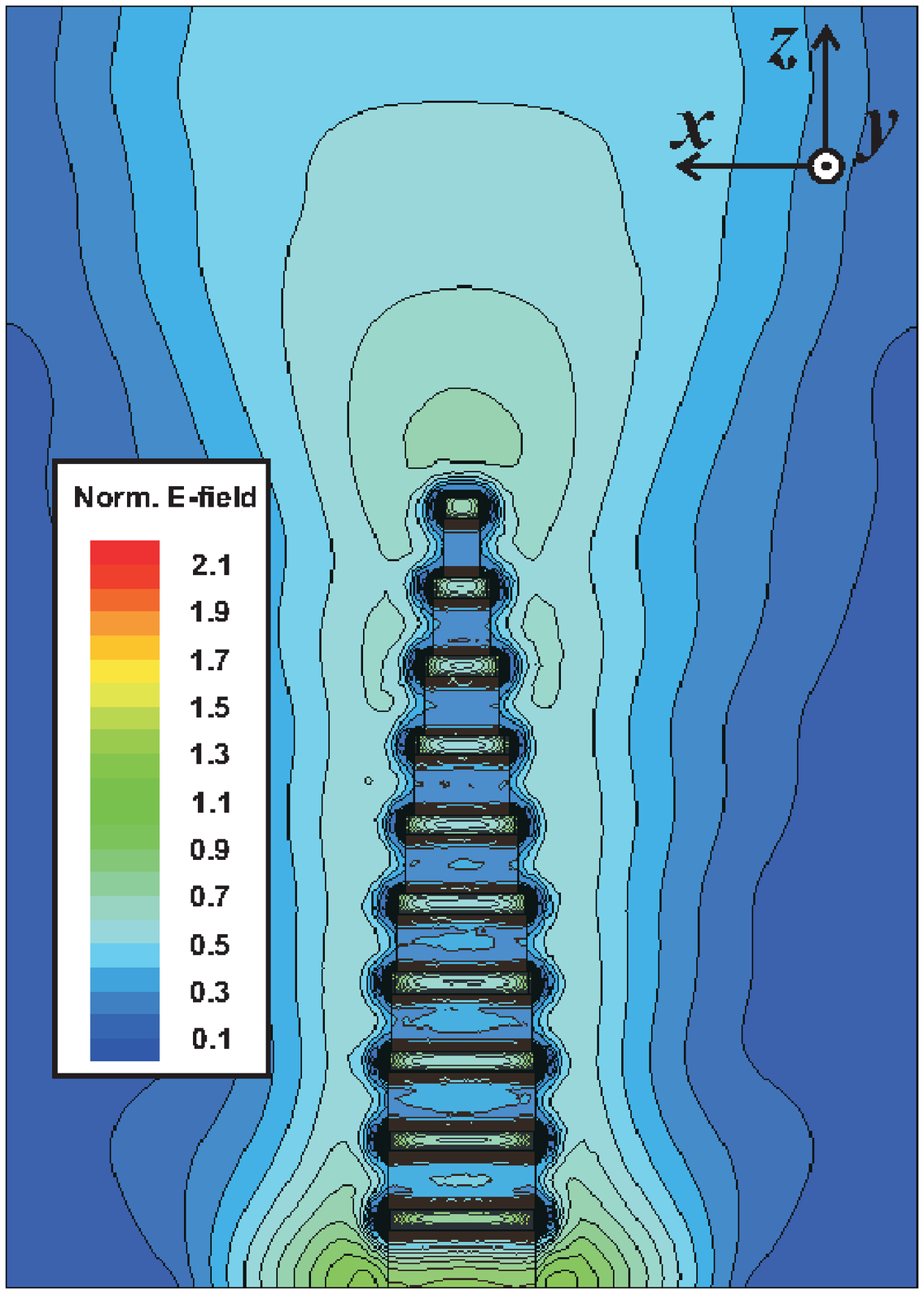,width=0.2\textwidth}}}
\mbox{
\subfigure[]{\epsfig{file=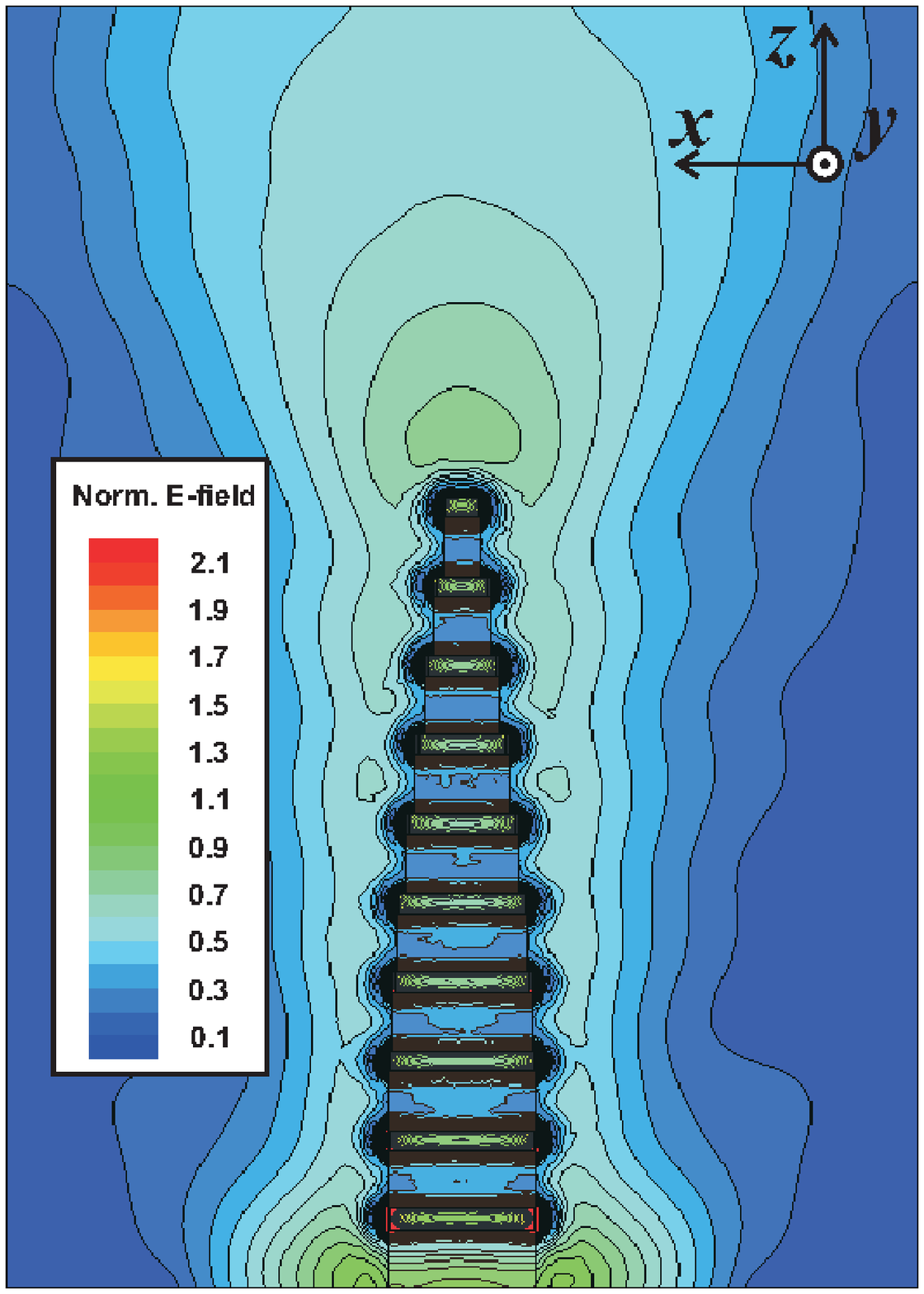,width=0.2\textwidth}}
\subfigure[]{\epsfig{file=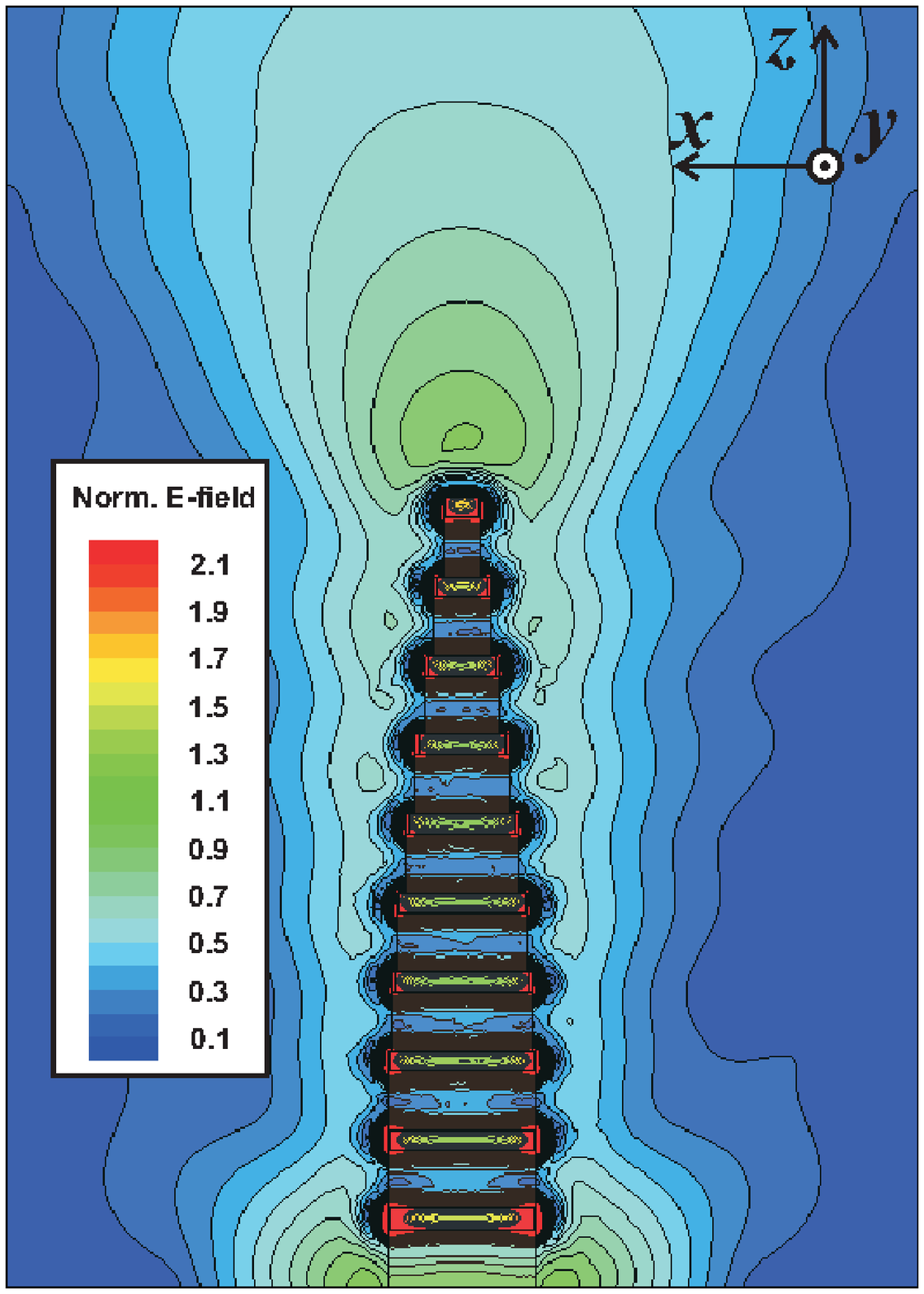,width=0.2\textwidth}}
\subfigure[]{\epsfig{file=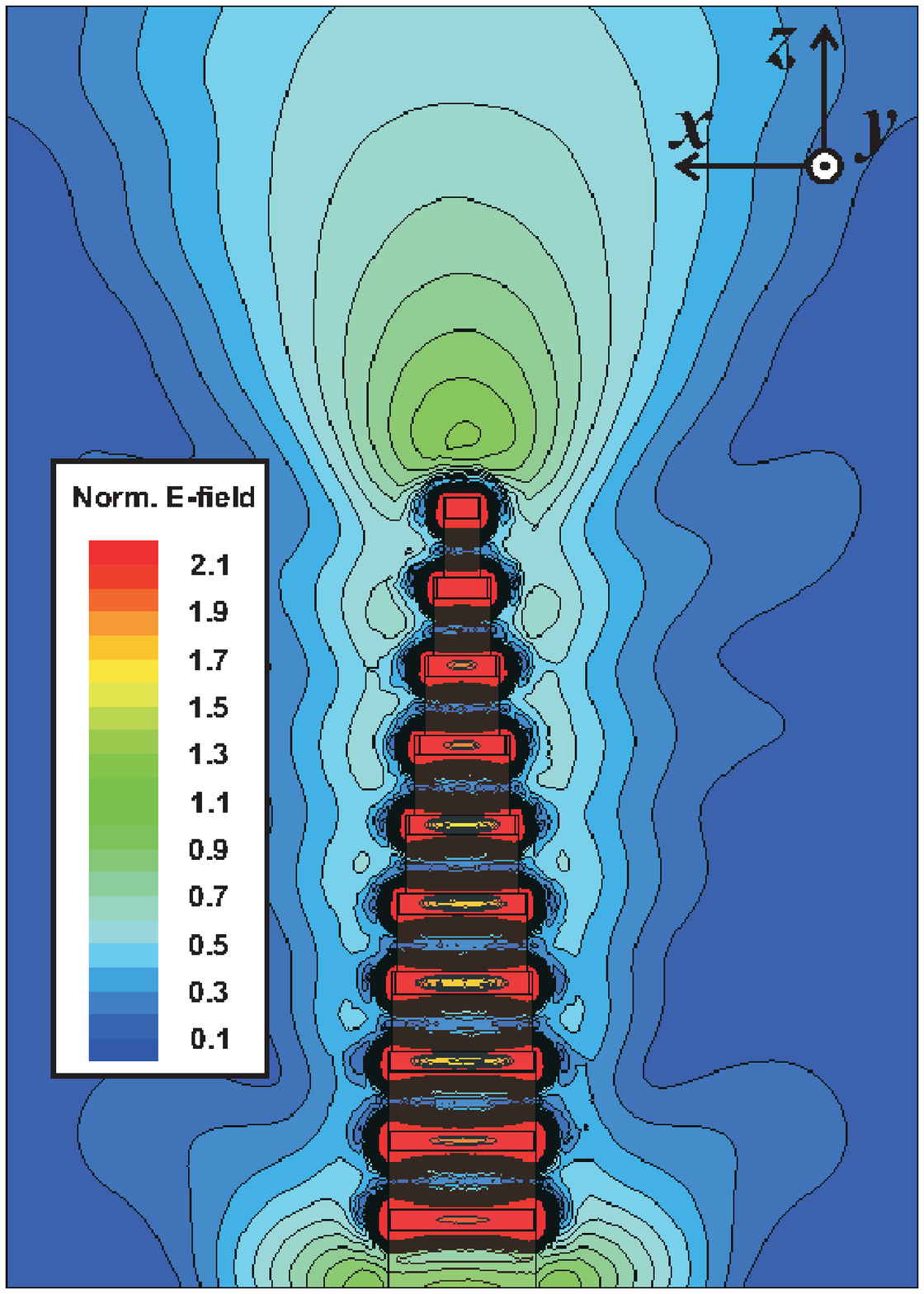,width=0.2\textwidth}}
\subfigure[]{\epsfig{file=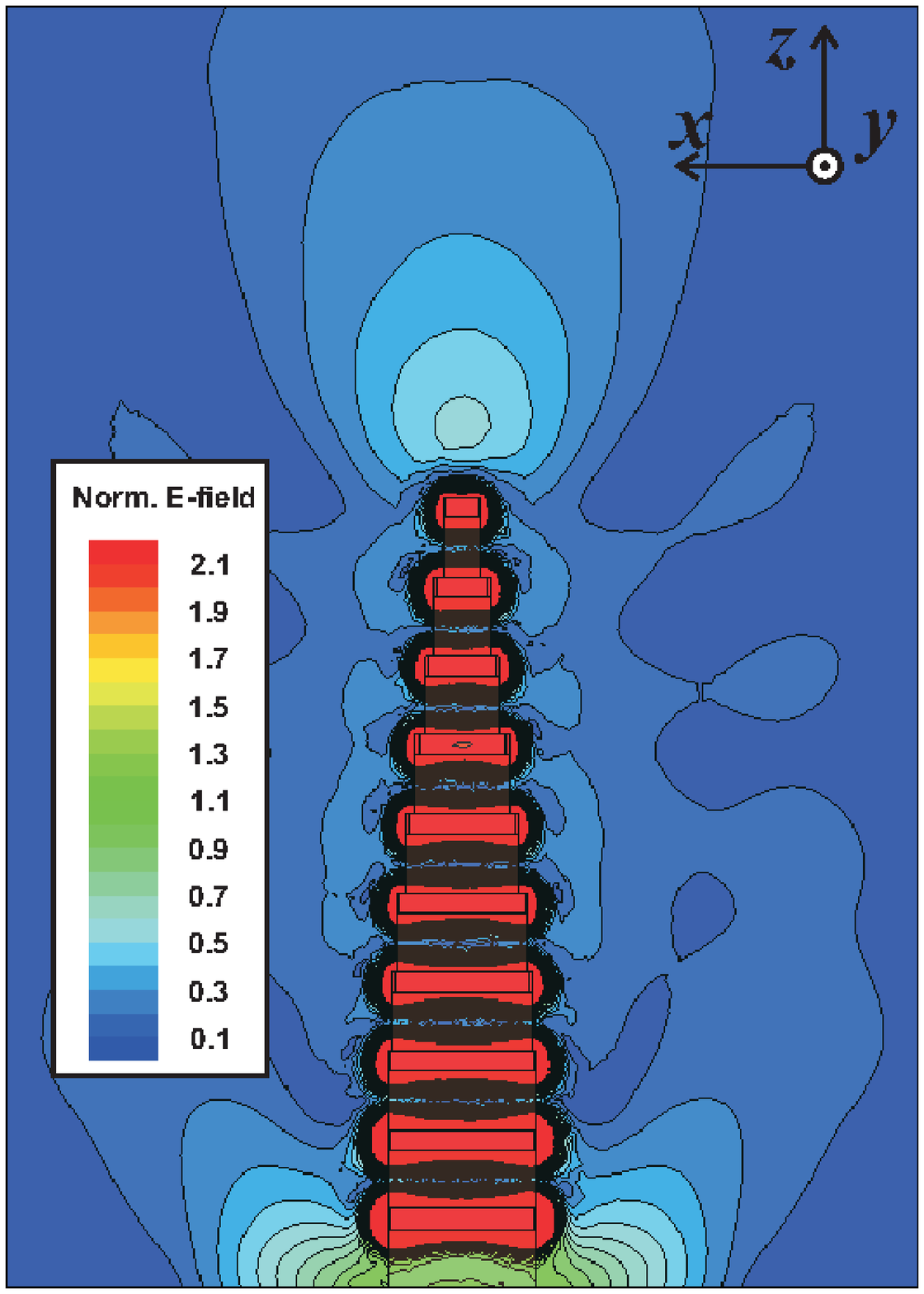,width=0.2\textwidth}}}
\caption{(Color online.) The normalized amplitude of the electric field at (a) 450\,THz, (b) 470\,THz, (c) 490\,THz, (d) 510\,THz, (e) 530\,THz, (f) 550\,THz, (g) 570\,THz, (h) 590\,THz, and (i) 610\,THz.}\label{nanotip_excitationport1}
\end{figure*}

We simulated the performance of the planar MMNT in HFSS. A
structure comprising 10 coplanar silver nanobars having thickness
$d=70$\,nm (dimension in $y-{\rm direction}$), length along the waveguide $d=35$\,nm (dimension in $z-{\rm
direction}$), and length across the waveguide ($x$-dim.) reducing
from $256$\,nm to $64$\,nm, {\itshape i.e.}, following the notations in Fig.~\ref{fig:1}, $a=32$\,nm and the length of last 6 nanoplates decreasing with step $b=32$\,nm. The structure was excited with a
wave port located at the input. The dielectric spacers had length in 
$z$-dimension ({\itshape i.e.}, $p-d$) $t=100$ nm and same dimensions in $x$- and $y$-direction as the silver nanobars.

In Figs.~\ref{nanotip_excitationport1}(a)--(i) we show the distribution of the absolute value of the electric
field amplitude normalized to that calculated in absence of the
plasmonic structure. The relative field distribution is depicted
for the central horizontal plane of the MMNT.

In all these figures the typical field distribution of the edge
wave is clearly seen in the narrow part of the MMNT. The field is
locally enhanced along the contour of the structure until the
distances of $30-70$ nm away from this contour. The amplitude
enhancement factor is equal $1.5\dots 7.5$ (depending on the
frequency). Within the frequency range $450-530$ THz the field
decreases fast at larger distances from the contour. Within the
frequency region $530-590$ THz the field is enhanced not only
along the contour. The constructive interference of two edge waves
observes at these frequencies and leads to a hot spot located near
the apex. The amplitude enhancement within this hot spot attains
$7\dots 8$ ({\itshape i.e.}, the intensity enhancement is $50\dots 60$).
Naturally, even a so large local field enhancement is not
comparable with that achievable in self-similar nanolenses.\cite{Stock} However, the combination of a rather large field
enhancement with a relatively large hot spot (its diameter is
$30\dots 50$ nm), a stable location of the hot spot in a very
broad frequency band makes our structure very attractive,
especially if we compare the fabrication costs of our structure to
that of a self-similar nanolens. At $590-610$ THz the field
enhancement along the contour disappears but the hot spot at the
apex keeps. At these frequencies the attenuation of the mode
becomes significant.

The main conclusion from Figs.~\ref{nanotip_excitationport1}
is that the regime which offers the local field enhancement in
front of the end of the MMNT refers to the whole frequency region
$450-610$ THz which covers $46$ percent of the visible range. In
this work we do not consider the matching of our nanostructure
with a conventional optical waveguide. This problem will be
addressed in our next paper.

\subsection{Pyramidal nanotip}

A systematic study of the pyramidal MMNT is not yet finished. In
the present paper we show results which will be hopefully improved
in the next future. Geometrical parameters selected below are
based only on analytical calculations. Simulations of the
nanopyramid were done using the FDTD codes.\cite{NT1,NT2,MMNT}
Two main differences in the operation of the pyramidal structure
from the planar one are as follows: 1) the multi-frequency
operation instead of the broadband one since the resonances of the
adjacent nanoplates do not overlap, 2) larger local enhancement in
the hot spot.

\begin{figure}[t!]
\centering
\epsfig{file=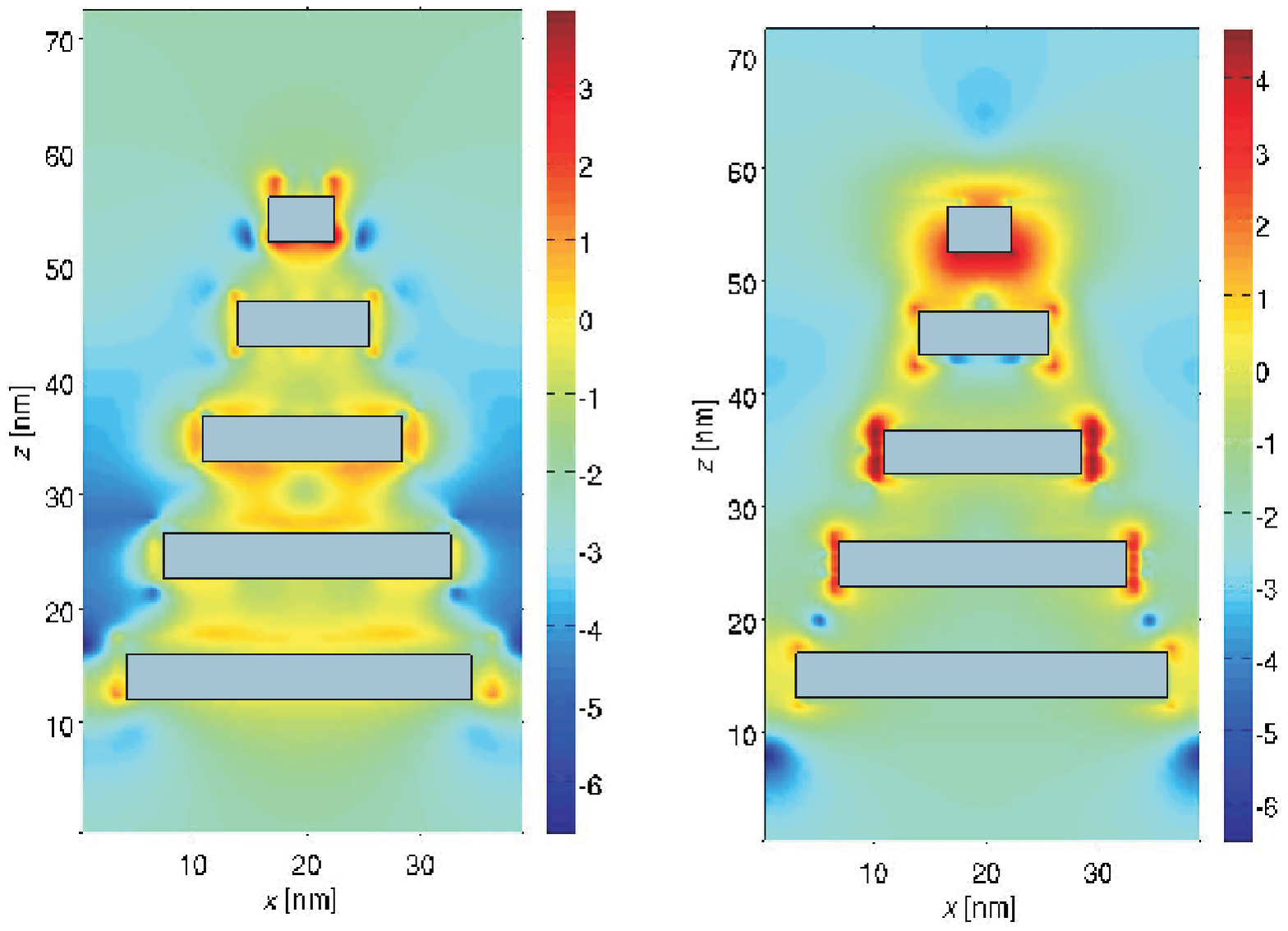,width=0.35\textwidth}
\epsfig{file=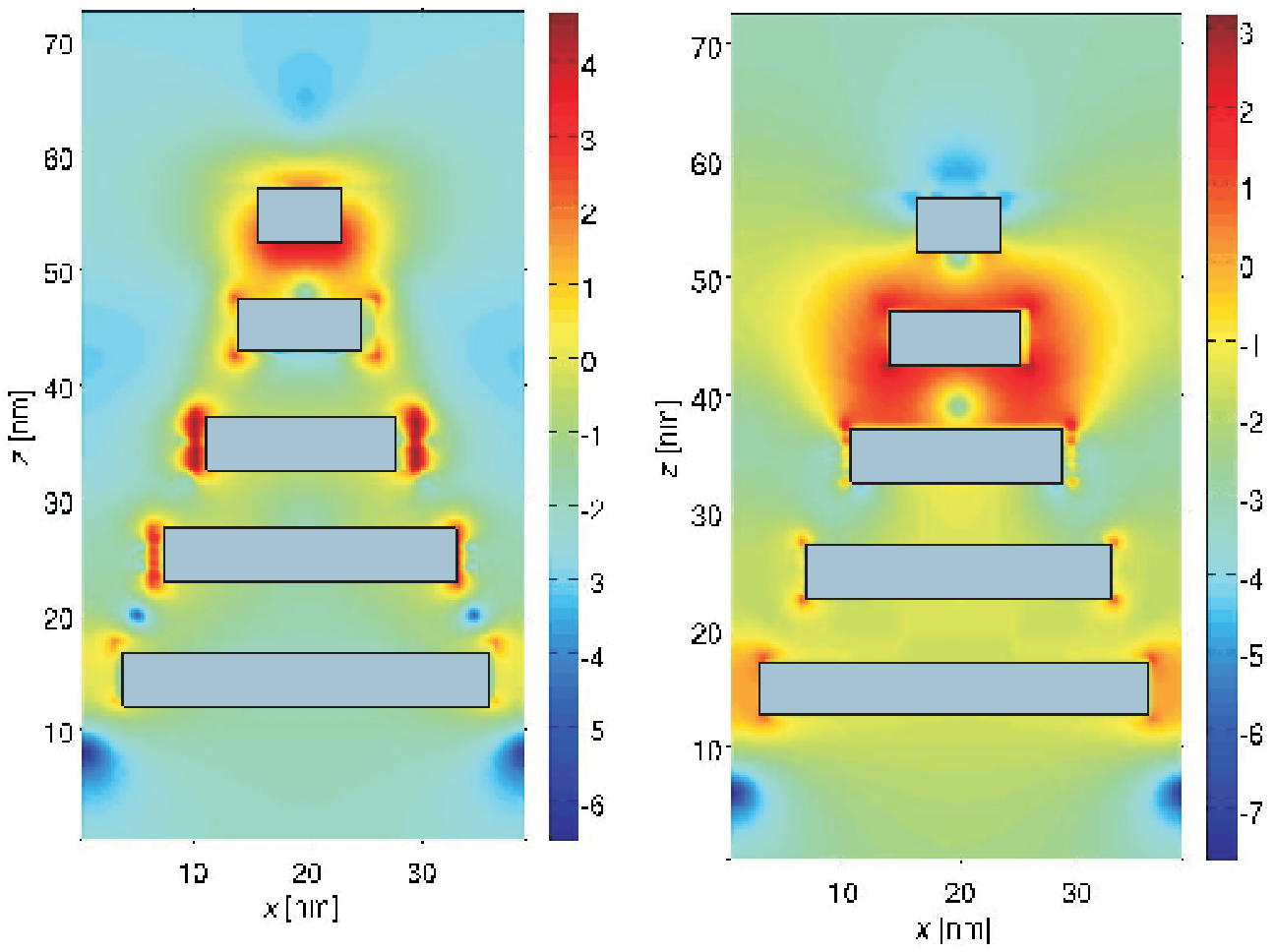,width=0.35\textwidth}
\epsfig{file=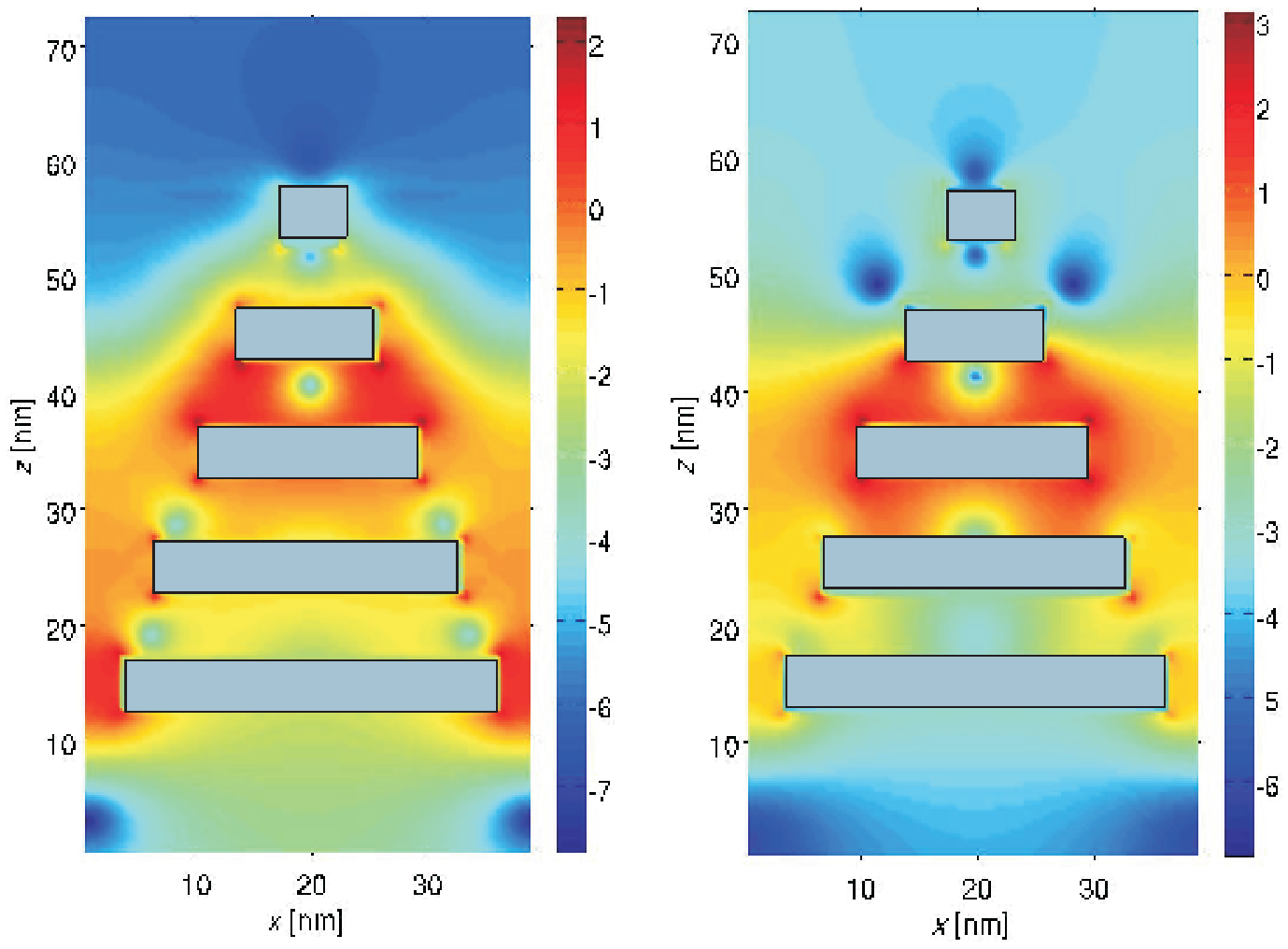,width=0.35\textwidth}
\caption{(Color online.) Logarithmic distribution of the field intensity over the
plane of axial cross section of the pyramidal nanostructure,
$\log(I/I_0)$. From upper left to down right the frequencies read: $862$\,THz, $680$\,THz, $590$\,THz, $526$\,THz, $462$\, THz, and $429$\,THz.} \label{Figa1}
\end{figure}

In simulations whose results are depicted in
Fig.~\ref{Figa1} the whole space around
the silver nanoplates was filled with uniform dielectric
$\varepsilon=2.2$. Thickness of silver nanoplates in this case was
$5$ nm as well as the thickness of dielectric spacings between the
plates. In Fig.~\ref{Figa1} we show
logarithmic field distributions $\log(I/I_0)$, where $I$ is the
field intensity in presence of nanoplates and $I_0$ is the field
intensity in their absence. The distributions are shown in the
plane of the pyramid axial cross section. This plane is parallel
to the plane in which the incident electric field is polarized.
The incident wave beam illuminates the pyramid base and is
directed along its axis.

The structure resonates at 9 frequencies ($385,\ 429,\ 462,\ 526,\
588,\ 632,\ 681,\ 769,\ 862$ THz). Each nanoplate has 2 plasmon
resonances except the smallest plate whose shape is close to cubic
and which has therefore one resonance. For five higher frequencies
we obtained the hot spot at the structure apex that we treat as
the result of the constructive interferences of edge waves in this
spatial region. For four lower resonance frequencies the
interference of edge waves  near the apex is destructive since the
field distribution has a minimum there. However at these four
frequencies we obtain nearly the same hot spot between the second
or third nanoplates (counted from the apex) because these
frequencies correspond to resonances of these two plates in two
orthogonal polarizations. So, with this structure we obtain one
rather stable hot spot for five discrete resonance frequencies and
another stable hot spot at other four frequencies. We also
simulated the structure with same $5$ nm dielectric spacing
between nanoplates but thickness of nanoplates was equal $2$ nm.
In such a structure similar situation observes with the location
of hot spots but the field enhancement in them is significantly
larger than that in Fig.~\ref{Figa1}.

\section{Conclusions}

In this paper we suggested, theoretically estimated and simulated
using the HFSS package and our own FDTD code new tapered
nanostructures which is our terminology refer to metamaterial
nanotips. These nanotips are formed by rectangular silver
nanoparticles, either nanoplates or nanobars alternating with
corresponding dielectric spacers which are arranged either in a
planar structure or in a pyramidal one. The structures are excited
by a wave beam incident from the structure bases. Subwavelength
spatial regions are formed around the structure and/or in front of
its apex in which the electric field is locally enhanced. In the
space surrounding the structure the field is significantly smaller
than the incident one. The advantage of our structure compared to
known ones is that the subwavelength spatial region of the
enhanced field is formed nearly the same at different frequencies.
For the planar structure the frequency stability holds over a very
broad frequency range. For the pyramidal one this is the stability
for several discrete frequencies. We believe that these
advantageous properties of our structures and their practical
feasibility will not only open new doors in field-enhanced schemes
of nanosensing and microscopy, they will be used in prospective
nanophotonic applications.

\section*{Acknowledgments}

This work was partially supported by a joint research grant of
Finnish Academy and Russian Foundation for Basic Research in the
field of magnetophotonics, plasmonics and nanooptics of
heterogeneous metamaterials. Authors are grateful to C.
Rockstuhl and S. M{\"u}hlig for useful discussion and auxiliary
simulations.

\end{document}